\documentclass[twocolumn,showpacs,preprintnumbers,amsmath,amssymb]{revtex4}

\usepackage[dvips]{graphicx}
\usepackage{dcolumn}
\usepackage{bm}

\begin{document}

\preprint{APS/123-QED}


\title{Soft-phonon-driven superconductivity in CaAlSi as seen by inelastic x-ray scattering}
\author{S. Kuroiwa, A. Q. R. Baron$^{1,2}$, T. Muranaka, R. Heid$^{3}$, K. -P. Bohnen$^{3}$}
\author{J. Akimitsu}%
\affiliation{
Department of Physics and Mathematics, Aoyama Gakuin University, Fuchinobe 5-10-1, Sagamihara, Kanagawa 229-8558, Japan\\
$^1$SPring-8$/$JASRI, 1-1-1 Kouto, Mikazuki-cho, Sayo-gun, Hyogo, 679-5198, Japan\\
$^2$SPring-8$/$RIKEN, 1-1-1 Kouto, Mikazuki-cho, Sayo-gun, Hyogo, 679-5148, Japan\\
$^3$Forshcungszentrum Karlsruhe, Institut f$\ddot{u}$r Festk$\ddot{o}$rperpysik, P. O. B. D-76021, Germany
}

\begin{abstract}
Inelastic x-ray scattering and $ab$-$initio$ calculation are applied to investigate the lattice dynamics and electron-phonon coupling of the ternary silicide superconductor CaAlSi ($P\bar{6}m2$).  
A soft $c$-axis polarized mode is clearly observed along the $\Gamma$-$A$-$L$ symmetry directions.  
The soft mode is strongly anharmonically broadened at room temperature, but, at 10 K, its linewidth narrows and becomes in good agreement with calculations of linear electron-phonon coupling. 
This establishes a coherent description of the detailed phonon properties in this system and links them clearly and consistently with the superconductivity.  
\end{abstract}

\maketitle

The discovery of superconductivity in MgB$_2$ \cite{nagamatsu} has refocused interest on electron-phonon coupling (EPC) as a basis for developing new superconductors with higher superconducting transition temperatures ($T_{\rm c}$).
For MgB$_2$, the high frequency optical $E_{\rm 2g}$ mode (B-B bond stretching) is strongly coupled with the $\sigma$ sheets of the Fermi surface \cite{Baron:MgB2,Shukla:MgB2}, while the average EPC for the full Fermi surface is moderate \cite{Choi}.
Such an anisotropic electron-phonon interaction enhances $T_{\rm c}$ in MgB$_2$ without electron-electron correlations.
Analogous intermetallic compounds with relatively high-$T_{\rm c}$ and honeycomb layered structure such as alkaline earth-intercalated graphites \cite{Weller} and ternary silicide $M$AlSi ($M$ = Ca, Sr, Ba) \cite{Imai,Imai0} are interesting to gain insight into the electron-phonon pairing mechanism. 
\par
Here we investigate phonons and their relation to superconductivity in CaAlSi, which has a structure similar to MgB$_2$.
However, whereas superconductivity in both materials is phonon mediated, and even, in both cases, is driven by specific optical phonon modes, in fact the detailed mechanisms are very different.
For MgB$_2$ a high energy phonon mode is strongly softened and broadened (linewidths $>$ 10 meV) by its coupling to the electronic system.
The phonon softening then is indicative of the coupling that drives the superconductivity.
In contrast, the superconductivity in CaAlSi is driven by structural instability leading to the presence of a soft mode related to out-of-plane Al$/$Si vibrations \cite{Huang,mazin,Matteo,Rolf2}.
The low frequency of the soft mode enhances its coupling to the electronic system and leads to a relatively high-$T_{\rm c}$.   In this paper we demonstrate the over-all picture of soft-mode driven superconductivity in CaAlSi is correct by measuring the properties of the soft mode throughout much of the Brillouin zone.  
This provides a nice contrast to work on MgB$_2$.  
\par
CaAlSi has several isomeric structures, as is unsurprising for a material near a structural instability.  
The common feature is alternating planes of Ca and planes of AlSi, but the stacking and flatness of the AlSi planes depends on detailed growth conditions \cite{Kuroiwa}: periodic structures in both the AlSi stacking and in the AlSi plane buckling are possible corresponding to 5 and 6 plane periodicity (5$H$- and 6$H$-CaAlSi, respectively) \cite{Sagayama}.  
More recently a simpler structure (1$H$-CaAlSi, see inset to fig. 1) without a $c$-axis superperiod and with flat AlSi planes has been made \cite{Kuroiwa}. 
This latter structure has $P\bar{6}m2$ symmetry, is similar to MgB$_2$, and is the subject of the present investigation.  
One notes that the buckling of the planes in the 5$H$ and 6$H$ structures follows the eigenvector of the soft mode discussed below. 

\par
We have investigated the phonon structure of 1$H$-CaAlSi using inelastic x-ray scattering (IXS) at BL35XU of SPring-8 \cite{Baron}.
The incident synchrotron x-ray beam at 21.747 keV was monochromatized by a Si (11 11 11) backscattering reflection providing $\sim$ 4 $\times$ 10$^9$ photons$/$s.
The scattered x-rays were collected using a two-dimensional array of 12 spherical analyzer crystals on the 10 m two-theta arm (horizontal scattering plane).
This allowed simultaneous measurement of several momentum transfers for longitudinal and transverse symmetries (using a horizontal and vertical line of analyzers, respectively), greatly facilitating data collection \cite{baron_array}.
The momentum resolution was set by slits to be typically $\sim$0.05 {\AA$^{-1}$} $\times$ 0.03 {\AA$^{-1}$ } (in and out of the scattering plane).
The energy resolution was determined from measurements of plexiglass to be 1.6 to 1.9 meV full width at half maximum (FWHM), depending on the analyzer crystal.  
\par
The sample was the same single crystal of 1$H$-CaAlSi (4 $\times$ 2 $\times$ 1 mm$^3$) used in previous studies \cite{Kuroiwa}.
The typical x-ray rocking curve of the single crystal was determined to be 0.07$^{\rm \circ}$ and 0.23$^{\rm \circ}$ for (2 2 0) and (0 0 4) reflections, respectively.
The $T_{\rm c}$ of 1$H$-CaAlSi was estimated to be 6.5(1) K by magnetic susceptibility, electrical resistivity and heat capacity measurements, with more details about its superconducting properties available in Ref. \cite{Kuroiwa}.
Crystal structure analysis using synchrotron x-ray and neutron diffraction technique suggested that 1$H$-CaAlSi has the space group $P\bar{6}m2$ with flat (un-buckled) AlSi planes and no $c$-axis superstructure, and its lattice constants were estimated to be approximately $a$ = 4.196(3) {\AA} and $c$ = 4.414(4) {\AA}.
Thus, the present sample has a hexagonal-reciprocal unit cell with $M$ point at 0.86 {\AA}$^{-1}$ and $A$ point at 0.71 {\AA}$^{-1}$. 
At zone center one can then expect 6 optical modes, two doubly degenerate $E'$ modes (in-plane motion) and two nondegenerate $A''_2$ modes (Al and Si displaced along the $z$-direction).

\par
Calculations of the lattice dynamics and EPC were performed in the framework of density functional theory using the linear response technique in combination with the mixed-basis pseudopotential method \cite{Rolf}.
The local-density approximation (LDA) was used for the exchange correlation potential as given by Hedin-Lundqvist \cite{Hiden}.
The optimized lattice constants are estimated to be $a$ = 4.155 {\AA} and $c$ = 4.269 {\AA}, slightly smaller than the experimental ones due to the usual LDA-overbinding effect.

\begin{figure}
\includegraphics[width=1\linewidth]{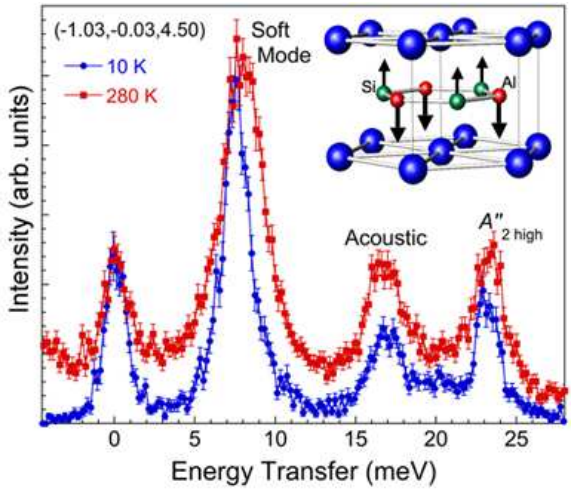}
\caption{\label{Fig1} (Color online) IXS spectra at 10 K (blue, circle) and 280 K (red, square) in almost longitudinal geometry measured at $\bf Q$ = (-1.03 -0.03 4.50) corresponding to $A$ point in the hexagonal Brillouin zone. Inset shows the eigenpolarization for the soft phonon mode.}
\end{figure}

\begin{figure}[!]
\includegraphics[width=1\linewidth]{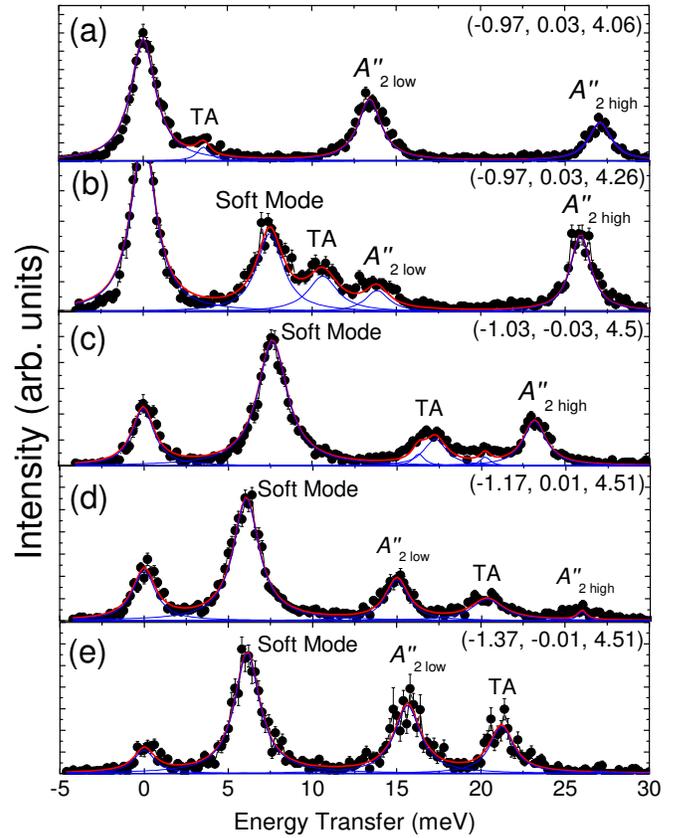}
\caption{\label{Fig2} (Color online) IXS spectra at 10 K along the $\Gamma$-$A$-$L$ lines. Panel (a) to (c) are data collected at $\bf Q$ = (-1 0 4+$\xi$), while (c) to (e) are along $A$-$L$ direction measured at $\bf Q$ = (-1+$\xi$ 0 4.5) [$\xi$ is expressed in reciprocal lattice units].  The solid line is a fit (see the text).}
\end{figure}

\par
Figure {\ref{Fig1}} shows representative IXS spectra at 10 K and 280 K at $\bf Q$ = (-1.03 -0.03 4.50) corresponding to the Brillouin zone boundary ($A$ point).
This geometry selects phonon modes with $c$-axis motion.
The soft phonon mode (``Soft Mode'') is readily apparent between 5 and 10 meV, in addition to one acoustic phonon and optical $A''_{\rm 2~high}$ mode (some small contribution from other modes is also expected near 16 and 20 meV).
The increased background at high temperatures is probably the result of multi-phonon scattering.  We estimate the Debye-Waller factors for the atoms ($e^{-2W}$) to be 0.68, 0.46 and 0.28 for Ca, Si and Al, respectively, at 300 K, so that a large multi-phonon contribution can be expected.

\par
As is clear from figure 1, the soft phonon mode exhibits narrowing and a frequency shift to lower energy at low temperatures, while other phonon modes show no significant change.
Its linewidth changes from 3.3(1) meV at 280 K to 2.0(1) meV at 10 K, and it softens from 7.9(1) meV to 7.5(1) meV.  
The narrowing is indicative of anharmonic effects near room temperature, while the softening at low temperature is a bit unusual, as generally phonons harden at low temperatures due to lattice contraction \cite{Kuro:lattice}. 
However, it might be accounted for by a quartic anharmonicity.  
The phonon energy can be generally described by $\omega^2$ = $\omega_{0}^2$ + $k_4\langle{u^2}\rangle$, where $\omega_0$ is the harmonic phonon energy, $k_4$ the coefficient of quartic anharmonic potential and $\langle{u^2}\rangle$ the square of anharmonic thermal displacement \cite{Stevenson}. 

Figure {\ref{Fig2}} shows IXS spectra at 10 K along the $\Gamma$-$A$-$L$ lines. 
To extract quantitative results, a least-squares fitting analysis was performed for the data at 10 K  using Lorentzian functions for the phonon modes.  
The mode intensities were fixed from calculated dynamical structure factor while the mode frequencies and linewidth were free.   
The soft phonon mode is slightly broader than the experimental resolution.
Meanwhile, we confirm that FWHM of $A''_{\rm 2~high}$ mode at 10 K is nearly equal to the experimental resolution at measured Q positions (but it very slightly broader).
In the absence of steep dispersion, broadening of the phonon linewidth beyond the experimental resolution corresponds to a reduced lifetime, and assuming negligible anharmonic contribution, can be directly related to the EPC.
Therefore, the soft mode is expected to have a larger EPC constant than that of other modes (which we discuss below in more detail).

\begin{figure}[t]
\includegraphics[width=1\linewidth]{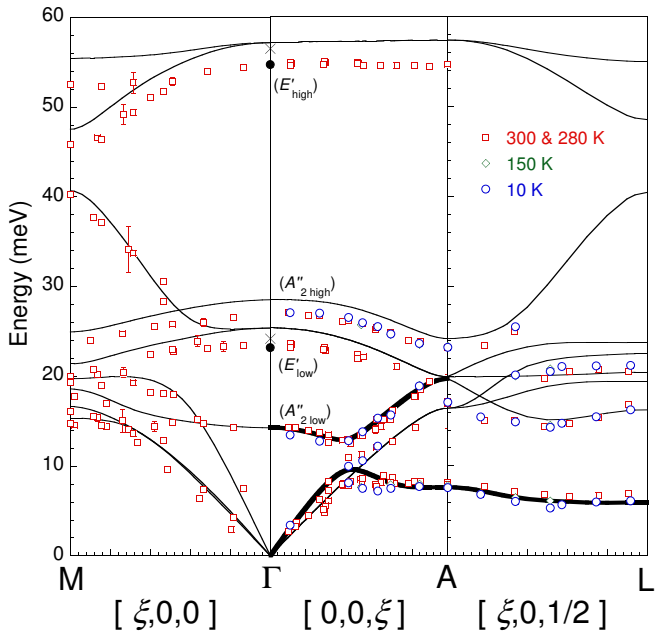}
\caption{\label{Fig3} (Color online) Phonon dispersion obtained by both IXS experiments (open symbols) and $ab$-$initio$ calculation (lines) of 1$H$-CaAlSi along $\Gamma$-$M$ and $\Gamma$-$A$-$L$ lines. The additional points at $\Gamma$ (filled circles and crosses) are data obtained by Raman scattering \cite{Kuroiwa:Raman,footnote:raman}.}
\end{figure}

Figure {\ref{Fig3}} shows phonon dispersion obtained by both IXS experiments (points) and $ab$-$initio$ calculation (lines).  There is good agreement with theory for the dispersion of the soft, acoustic and optical $A''_{\rm 2~low}$ branches.
On the other hand, differences appear in the $A''_{\rm 2~high}$ and $E'$ phonon branches, with the measured results being at lower frequency than the calculations.  This probably is the result of the tendency of the LDA to underestimate the equilibrium volume.
The present calculation produces an equilibrium volume 5 $\%$ smaller than the measured value and stiffens force constants, which leads to the frequencies of the optical branches to be higher by up to 3-5 $\%$ of the experimental data.
\par
The dispersion of the soft mode is in excellent agreement with calculation throughout zone.
An anticrossing between the optical $A''_{\rm 2~low}$ and soft/acoustic mode at around 0.4 $\Gamma$-$A$ was clearly confirmed by our data, in agreement with the present calculation but not others {\cite{Huang,Matteo}}.
Moreover, the frequency shift manifested in the low temperature measurements was also observed in the $A''_{\rm 2~low}$ mode around the zone center, suggesting some anharmonicity. 
Meanwhile, the energy of both $E'$ modes near the zone center obtained from IXS is identical to those with lower energy in each Raman data as shown in Fig. {\ref{Fig3}}.

\begin{figure}[t]
\includegraphics[width=1\linewidth]{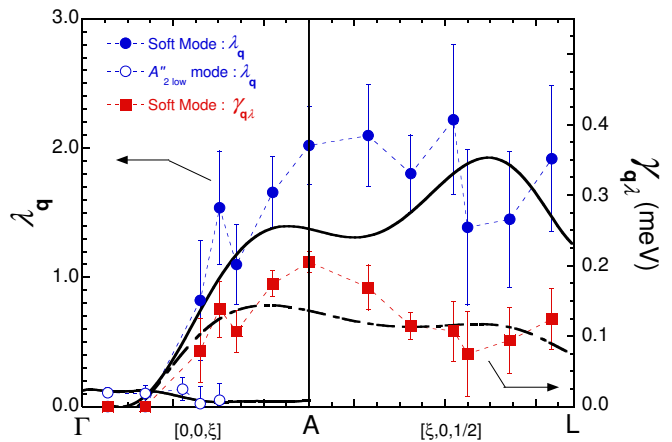}
\caption{\label{Fig4} (Color online) Experimental (at 10 K, symbols) and calculated (solid line) electron-phonon coupling constant $\lambda_{\bf q}$ of soft mode and $A''_{\rm 2~low}$ modes along the $\Gamma$-$A$-$L$ lines. Square symbols and dot-dashed line correspond to the measured and calculated data of electronic contribution to the linewidth for the soft mode, respectively. The calculated results correspond to those of bold-lines in Fig. {\ref{Fig3}}.}
\end{figure}

We now discuss the relationship between the electron-phonon interaction of the soft phonon mode and the superconductivity in 1$H$-CaAlSi based on the measured linewidth at the low temperature.
The mode-resolved EPC mass enhancement factor $\lambda_{\bf q{\lambda}}$ is obtained from the phonon spectral linewidth and frequency using \cite{Allen1,Allen2},
\begin{equation}
\lambda_{\bf q{\lambda}}=\frac{2}{\pi{N(0)}}\frac{\gamma_{\bf q{\lambda}}}{\omega^2_{\bf q{\lambda}}}  \label{ep}
\end{equation}
where $N(0)$ is the electronic density of states at the Fermi energy, $\omega_{\bf q{\lambda}}$ is the phonon frequency and $\gamma_{\bf q{\lambda}}$ is the electronic contribution to the linewidth (half-width at half-maximum) of a phonon mode (${\bf q{\lambda}}$). 
Since the $N(0)$ is estimated to be 1.10 (states/eV unitcell) from electronic structure calculation using linearized augmented planewave method \cite{Kuroiwa:Band}, we can experimentally determine the EPC constant $\lambda_{\bf q}$ for the soft mode and $A''_{\rm 2~low}$ low modes using eq. (1). 
As seen in Fig \ref{Fig4}, the agreement between the measured (symbols) and calculated (lines) $\lambda_{\bf q}$ is good.  
Other optical phonon modes have a smaller $\lambda_{\bf q{\lambda}}$ in comparison to the soft phonon mode.   Thus these results provide strong experimental evidence supporting soft-phonon-driven superconductivity in 1$H$-CaAlSi.

In summary, we have investigated the lattice dynamical properties and electron-phonon coupling in 1$H$-CaAlSi by combined inelastic x-ray scattering experiments and $ab$-$initio$ calculations. 
Overall measurements generally show excellent agreement with mixed basis LDA pseudopotential calculations. 
The soft phonon mode, consisting of out-of-plane Al$/$Si vibration was readily apparent along the 0.4 $\Gamma$-$A$ to $A$-$L$ path, showing both large anharmonicity and strong electron-phonon coupling. 
Thus, we established a coherent description of the correlation between the detailed phonon properties and superconductivity of 1$H$-CaAlSi from viewpoint of both experiments and theory.

\vspace{3mm}

This work was carried out at SPring-8 under proposal No. 2006B1082 and 2007A1523. We acknowledge financial support by the 21 st COE program, "High-Tech Research Center" Project for Private Universities: matching fund subsidy from MEXT (Ministry of Education, Culture, Sports, Science and Technology; 2002-2004) and the Grant-in-Aid for Japan Society for the Promotion of Science (JSPS) Fellows.


\begin{thebibliography}{99}

\bibitem{nagamatsu} J. Nagamatsu, N. Nakagawa, T. Muranaka, Y. Zenitani, and J. Akimitsu, Nature (London) {\bf 410}, 63 (2001).
\bibitem{Baron:MgB2} A. Q. R. Baron, H. Uchiyama, Y. Tanaka, S. Tsutsui, D. Ishikawa, S. Lee, R. Heid, K. -P. Bohnen, S. Tajima, and T. Ishikawa, Phys. Rev. Lett. {\bf 92}, 197004 (2004).
\bibitem{Shukla:MgB2} A. Shukla, M. Calandra, M. d'Astuto, M. Lazzeri, F. Mauri, C. Bellin, M. Krisch, J. Karpinski, S. M. Kazakov, J. Jun, D. Daghero, and K. Parlinski, Phys. Rev. Lett. {\bf 90}, 095506 (2003).
\bibitem{Choi} H. J. Choi, D. Roundy, H. Sun, M. L. Cohen, and S. G. Louie, Phys. Rev. B {\bf 66}, 020513 (2002).
\bibitem{Weller} T. E. Weller, M. Ellerby, S. S. Saxena, R. P. Smith, and N. T. Skipper, Nat. Phys. {\bf 1}, 39 (2005).
\bibitem{Imai0}M. Imai, E Abe, J. Ye, K. Nishida, T. Kimura, K. Honma, H. Abe, and H. Kitazawa, Phys. Rev. Lett.  {\bf 87}, 077003 (2001).
\bibitem{Imai} M. Imai, E. S. Sadki, H. Abe, K. Nishida, T. Kimura T. Sato, K Hirata, and H. Kitazawa, Phys. Rev. B {\bf 68}, 064512 (2003).
\bibitem{Huang} G. Q. Huang, L. F. Chen, M. Liu, and D. Y. Xing, Phys. Rev. B {\bf 71}, 172506 (2005).
\bibitem{mazin} I. I. Mazin and  D. A. Papaconstantopoulos, Phys. Rev. B {\bf 69}, 180512(R) (2004).
\bibitem{Matteo} M. Giantomassi, L. Boeri, and G. B. Bachelet, Phys. Rev. B {\bf 72}, 224512 (2005).
\bibitem{Rolf2} R. Heid, K. -P. Bohnen, B. Renker, P. Adelmann, T. Wolf, D. Ernst, and H. Schober, J. Low Temp. Phys. {\bf 147}, 375 (2007).
\bibitem{Kuroiwa} S. Kuroiwa, H. Sagayama, T. Kakiuchi, H. Sawa, Y. Noda, and J. Akimitsu, Phys. Rev. B {\bf 74}, 014517 (2006).
\bibitem{Sagayama} H. Sagayama, Y. Wakabayashi, H. Sawa, T. Kamiyama, A. Hoshikawa, S. Harjo, K. Uozato, A. K. Ghosh, M. Tokunaga, and T. Tamegai, J. Phys. Soc. Jpn. {\bf 75}, 043713 (2006).
\bibitem{Baron} A. Q. R. Baron, Y. Tanaka, S. Goto, K. Takeshita, T. Matsushita, and T. Ishikawa, J. Phys. Chem. Solids {\bf 61}, 461 (2000).
\bibitem{baron_array} A. Q. R. Baron, $el$ $al$., J. Phys. Chem. Solids, Accepted.
\bibitem{Rolf} R. Heid and K. -P. Bohnen, Phys. Rev. B {\bf 60}, 3709(R) (1999).
\bibitem{Hiden} L. Hiden and B. J. Lundqvist, J. Phys. C {\bf 4}, 2064 (1971).
\bibitem{Kuro:lattice}  The lattice expansion was measured to be $\Delta{c}/c$ $\sim$ 7.2 $\times$ 10$^{-3}$ and $\Delta{a}/a$ $\sim$ 7.1 $\times$ 10$^{-4}$ between 20 K and 300 K.
\bibitem{Stevenson} $Phonons$ $in$ $Perfect$ $Lattices$ $and$ $in$ $Lattices$ $with$ $Point$ $Imperfections$, edited by R.W. H. Stevenson (Plenum Press, New York, 1966).
\bibitem{Kuroiwa:Raman} S. Kuroiwa, T. Hasegawa, T. Kondo, N. Ogita, M. Udagawa, and J. Akimitsu (unpubished).
\bibitem{footnote:raman} Note that the higher energy Raman-active modes (crosses in fig. 3) were not observed, despite a careful search, in all IXS spectra, except one.  For the spectrum where a weak peak appeared, calculations suggest it is from a two-phonon contribution \cite{baron:twophonon}.
\bibitem{baron:twophonon} A. Q. R. Baron, H. Uchiyama, R. Heid, K. P. Bohnen, Y. Tanaka, S. Tsutsui, D. Ishikawa, S. Lee, and S. Tajima, Phys. Rev. B {\bf 75}, 020505(R) (2007).
\bibitem{Allen1} P. B. Allen and R. Silberglitt, Phys. Rev. B {\bf 9}, 4733 (1974).
\bibitem{Allen2} P. B. Allen, ``$Phonons$ $and$ $the$ $Superconducting$ $Transition$ $Temperature$'' 
in ``$Dynamical$ $Properties$ $of$ $Solids$'' edited by G. K. Horton and A. A. Maradudin, (North-Holland, Amsterdam, 1980) 95-196. Note for consistency with [23] $N(0)$ in eqn 4.27 should be taken as the total density of states.
\bibitem{Kuroiwa:Band} S. Kuroiwa, A. Nakashima, S. Miyahara, N. Furukawa, and J. Akimitsu, J. Phys. Soc. Jpn. {\bf 76}, 113705 (2007).

\end{thebibliography}
\end{document}